\documentclass[aps,preprint]{revtex4}%
\usepackage{amsfonts}
\usepackage{amsmath}
\usepackage{amssymb}
\usepackage{graphicx}%
\providecommand{\U}[1]{\protect \rule{.1in}{.1in}}

\begin{document}
\title{Polaronic properties of an impurity in a Bose-Einstein condensate in reduced dimensions.}
\author{W. Casteels$^{1}$, J. Tempere$^{1,2}$ and J. T. Devreese$^{1}$}
\affiliation{$^{1}$TQC, Universiteit Antwerpen, Groenenborgerlaan 171, B2020 Antwerpen, Belgium}
\affiliation{$^{2}$Lyman Laboratory of Physics, Harvard University, Cambridge,
Massachusetts 02138, USA}

\begin{abstract}
The application of optical lattices allows a tuning of the geometry of
Bose-Einstein condensates to effectively reduced dimensions. In the context of
solid state physics the consideration of the low-dimensional Fr\"{o}hlich
polaron results in an extension of the polaronic strong coupling regime. With
this motivation we apply the Jensen-Feynman variational principle to calculate
the ground state properties of the polaron consisting of an impurity in a
Bose-Einstein condensate in reduced dimensions. Also the response of this
system to Bragg scattering is calculated. We show that reducing the dimension
leads to a larger amplitude of the polaronic features and is expected to
facilitate the experimental observation of polaronic properties. In optical
lattices not only Feshbach resonances but also confinement-induced resonances
can be used to tune the polaronic coupling strength. This opens up the
possibility to experimentally reveal the intermediate and strong polaronic
coupling regimes and resolve outstanding theoretical questions regarding
polaron theory.

\end{abstract}
\maketitle

\section{Introduction}

In recent years ultracold atomic systems have revealed themselves as quantum
simulators for many-body theories \cite{RevModPhys.80.885}. Especially their
high degree of tunability makes them attractive for this purpose. An example
of a system that can be simulated in this way is the Fr\"{o}hlich polaron
which is well-known from solid state physics where it is used to describe
charge carriers in a polar solid (see for example Ref. \cite{BoekDevreese} for
an extended overview). In the context of ultracold gases the system of
impurities embedded in a Bose-Einstein condensation can be mapped onto the
Fr\"{o}hlich polaron Hamiltonian \cite{PhysRevLett.96.210401,
PhysRevA.73.063604}. In this case the role of the charge carriers is played by
the impurities and the lattice vibrations are replaced by the Bogoliubov
excitations. Recently this system has gained much interest both theoretically
\cite{PhysRevA.76.011605, 1367-2630-10-3-033015, 0295-5075-82-3-30004,
PhysRevA.82.063614, 1367-2630-13-10-103029, PhysRevB.80.184504,
springerlink:10.1134/S1054660X11150035, PhysRevA.84.063612,
PhysRevA.83.033631, springerlink:10.1007/s10909-010-0286-0}\ and
experimentally \cite{springerlink:10.1007/s00340-011-4868-6,
PhysRevLett.105.133202, PhysRevA.77.011603, PhysRevLett.105.045303,
2011arXiv1106.0828C}.

For the present work we focus on a single Fr\"{o}hlich polaron for which the
Hamiltonian can not be analytically diagonalized and one has to rely on
approximation methods. The most advanced theory for the ground state
properties is the Jensen-Feynman variational principle \cite{PhysRev.97.660}
which can be extended through the Feynman-Hellwarth-Iddings-Platzman (FHIP)
approximation for the response properties \cite{PhysRev.127.1004,
PhysRevB.5.2367}. The optical absorption of the Fr\"{o}hlich solid state
polaron was later also obtained through a diagrammatic Monte Carlo calculation
and a comparison with the FHIP approximation showed a good agreement at weak
and intermediate polaronic coupling but in the strong coupling regime
deviations were revealed \cite{PhysRevLett.91.236401, PhysRevLett.96.136405}.
Since there is no known material that exhibits the strong coupling behavior
only the weak and intermediate coupling regime could be experimentally probed
which resulted in a good agreement with the theory \cite{PolaronsAndExcitons,
PhysRevLett.58.1471}. A better understanding of the strong coupling regime
could also shed light on the possible role of polarons and bipolarons in
unconventional pairing mechanisms for high-temperature superconductivity
\cite{BoekAlexandrov, PhysRevB.77.094502}. Recently it was shown that for an
impurity in a condensate the use of a Feshbach resonance allows an external
tuning of the polaronic coupling parameter which makes it a promising system
to probe the strong polaronic coupling regime for the first time
\cite{PhysRevB.80.184504}. Recent experiments have shown the feasibility of
using Feshbach resonances for the tuning of interparticle interactions between
different species \cite{PhysRevA.85.042721, PhysRevA.85.032506,
PhysRevA.85.051602}.

Since the impurities are considered as not charged it is not possible to
conduct optical absorption measurements to reveal the polaronic excitation
structure as is possible for the Fr\"{o}hlich solid state polaron. It was
shown in \cite{PhysRevA.83.033631} that Bragg spectroscopy is suited to
experimentally probe the polaronic excitation structure of an impurity in a
condensate. Bragg scattering is a well established experimental technique in
the context of ultracold gases (see for example Refs.
\cite{PhysRevLett.88.120407, PhysRevLett.83.2876}). The setup consists of two
laser beams with different frequencies $\omega_{1}$ and $\omega_{2}$ and
different momenta $\vec{k}_{1}$ and $\vec{k}_{2}$ that are radiated on the
impurity. The impurity can then absorb a photon from laser 1 and emit it to
laser 2 during which process it has gained an energy $\omega=\omega_{1}%
-\omega_{2}$ and a momentum $\vec{k}=\vec{k}_{1}-\vec{k}_{2}$. The response is
reflected in the number of impurities that have gained a momentum $\vec{k}$ as
a function of $\vec{k}$ and $\omega$. This number is proportional to the
imaginary part of the density response function $\chi \left(  \omega,\vec
{k}\right)  $ \cite{Pitaevskii}:%

\begin{equation}
\chi \left(  \omega,\vec{k}\right)  =\frac{i}{\hbar}\int_{0}^{\infty
}dte^{i\omega t}\left \langle \left[  \widehat{\rho}_{\vec{k}}\left(  t\right)
,\widehat{\rho}_{\vec{k}}^{\dag}\right]  \right \rangle , \label{DensResp}%
\end{equation}
with $\widehat{\rho}_{\vec{k}}$ the density operator of the impurity.

Another powerful tool in the context of ultracold gases is the application of
optical lattices which can be employed to modify the geometry of the system
\cite{citeulike:2749190}. This allows to confine the system in one or two
directions such that the confinement length is much smaller than all other
typical length scales which results in an effectively low dimensional system.
For these systems the interparticle interactions can be described through a
contact pseudopotential with an amplitude that is a function of the
three-dimensional scattering length and the confinement length. This permits
to experimentally tune the interactions between the particles by varying the
strength of the confinement which results in a resonant behavior. These
confinement-induced resonances have been studied both theoretically
\cite{PhysRevLett.101.170401, 1367-2630-7-1-192, PhysRevLett.91.163201,
PhysRevLett.81.938, PhysRevA.64.012706} and experimentally
\cite{PhysRevLett.104.153203, PhysRevLett.94.210401, Haller04092009,
PhysRevLett.104.153202, PhysRevLett.106.105301}.

In the present work we adapt the calculations of the ground state and response
properties of the polaronic system consisting of an impurity in a condensate
to the case of reduced dimensions. This was done for the Fr\"{o}hlich solid
state polaron in Refs. \cite{PhysRevB.33.3926, PhysRevB.31.3420,
PhysRevB.36.4442}\ which led to the polaronic scaling relations. These are
applicable for polaronic systems of which the interaction amplitude
$V_{\vec{k}}$ (see later) is a homogeneous function. Unfortunately this is not
the case for the polaron consisting of an impurity in a Bose-Einstein
condensate. We start by showing that also in lower dimensions the Hamiltonian
of an impurity in a condensate can be mapped onto the Fr\"{o}hlich polaron
Hamiltonian. Then the Jensen-Feynman variational principle is applied to
calculate an upper bound for the free energy and an estimation of the
effective mass and the radius of the polaron as was done in Ref.
\cite{PhysRevB.80.184504}\ for the three-dimensional case. Subsequently the
treatment of Ref. \cite{PhysRevA.83.033631} for the response to Bragg
spectroscopy in 3 dimensions is adapted to reduced dimensions. All results are
applied to the specific system of a lithium-6 impurity in a sodium condensate.

\section{Impurity in a condensate in $d$ dimensions}

The Hamiltonian of an impurity in an interacting bosonic gas is given by:%
\begin{equation}
\widetilde{H}=\frac{\widehat{p}^{2}}{2m_{I}}+\sum_{\vec{k}}E_{\vec{k}}%
\widehat{a}_{\vec{k}}^{\dag}\widehat{a}_{\vec{k}}+\frac{1}{2}\sum_{\vec
{k},\vec{k}^{\prime},\vec{q}}V_{BB}\left(  \vec{q}\right)  \widehat{a}%
_{\vec{k}^{\prime}-\vec{q}}^{\dag}\widehat{a}_{\vec{k}+\vec{q}}^{\dag}%
\widehat{a}_{\vec{k}}\widehat{a}_{\vec{k}^{\prime}}+\sum_{\vec{k},\vec{q}%
}V_{IB}\left(  \vec{q}\right)  e^{i\vec{q}.\widehat{r}}\widehat{a}_{\vec
{k}^{\prime}-\vec{q}}^{\dag}\widehat{a}_{\vec{k}^{\prime}}. \label{HamOrig}%
\end{equation}
The first term in this expression represent the kinetic energy of the impurity
with $\widehat{p}$ ($\widehat{r}$) the momentum (position) operator of the
impurity with mass $m_{I}$. The second term in the right-hand side of
(\ref{HamOrig}) describes the kinetic energy of the bosons with creation
(annihilation) operators $\left \{  \widehat{a}_{\vec{k}}^{\dag}\right \}  $
($\left \{  \widehat{a}_{\vec{k}}\right \}  $) and energy $E_{\vec{k}}%
=\frac{\hbar^{2}k^{2}}{2m_{B}}-\mu$ where $\mu$ is the chemical potential of
the bosons and $m_{B}$ their mass. The last two terms represent the
interaction energy with $V_{BB}\left(  \vec{q}\right)  $ the Fourier transform
of the boson-boson interaction potential and $V_{IB}\left(  \vec{q}\right)  $
of the impurity-boson interaction potential. All vectors in expression
(\ref{HamOrig}) are considered as $d$-dimensional.

In Refs. \cite{PhysRevLett.85.3745} and \cite{PhysRevLett.84.2551} it is shown
that in one and two dimensions, respectively, at temperatures well below a
critical temperature $T_{c}$ a trapped weakly interacting Bose gas is
characterized by the presence of a true condensate while just below $T_{c}$
this is a quasicondensate. A quasicondensate exhibits phase fluctuations with
a radius $R_{\phi}$ that is smaller than the size of the system but greatly
exceeds the coherence length $\xi$ \cite{PhysRevLett.85.3745,
PhysRevLett.84.2551}. Since the radius of the polaron $R_{pol}$ is typically
of the order $\xi$ (see later) we have $R_{pol}\ll R_{\phi}$ which shows that
the polaronic features are also present in a quasicondensate. In the following
we no longer make the distinction and use the name condensate for both
situations. The presence of a condensate can be expressed through the
Bogoliubov shift which (within the local density aproximation) transforms the
Hamiltonian (\ref{HamOrig}) into \cite{PhysRevB.80.184504}:%
\begin{equation}
\widehat{H}=E_{GP}+g_{IB}N_{0}+\widehat{H}_{pol},\label{TotHamilt}%
\end{equation}
where use was made of contact interactions, i.e. $V_{BB}\left(  \vec
{q}\right)  =g_{BB}$ and $V_{IB}\left(  \vec{q}\right)  =g_{IB}$. In order to
have a stable condensate the boson-boson interaction should be repulsive, i.e.
$g_{BB}>0$. The sign of the impurity-boson interaction strength $g_{IB}$ is in
priciple arbitrary, however for the Bogoliubov approximation to be valid the
depletion of the condensate around the impurity must remain smaller than the
condensate density which means the formalism is not valid for a large negative
$g_{IB}$ \cite{0295-5075-82-3-30004, PhysRevLett.102.030408}. The first term
in the right-hand side of (\ref{TotHamilt}) represents the Gross Pitaevskii
energy $E_{GP}$ of the condensate and the second term gives the interaction of
the impurity with the condensate (with $N_{0}$ the number of condensed bosons
in a unit volume). The third term is the polaron Hamiltonian which describes
the interaction between the impurity and the Bogoliubov excitations:%
\begin{equation}
\widehat{H}_{pol}=\frac{\widehat{p}^{2}}{2m_{I}}+\sum_{\vec{k}\neq0}%
\hbar \omega_{\vec{k}}\widehat{\alpha}_{\vec{k}}^{\dag}\widehat{\alpha}%
_{\vec{k}}+\sum_{\vec{k}\neq0}V_{\vec{k}}\rho_{I}\left(  \vec{k}\right)
\left(  \widehat{\alpha}_{\vec{k}}+\widehat{\alpha}_{-\vec{k}}^{\dag}\right)
,\label{PolHam}%
\end{equation}
where $\left \{  \widehat{\alpha}_{\vec{k}}^{\dag}\right \}  $ ($\left \{
\widehat{\alpha}_{\vec{k}}\right \}  $) are the creation (annihilation)
operators of the Bogoliubov excitations with dispersion:%
\begin{equation}
\hbar \omega_{\vec{k}}=\frac{\hbar^{2}k}{2m_{B}\xi}\sqrt{\left(  \xi k\right)
^{2}+2},\label{BogSpectr}%
\end{equation}
with $\xi$ the healing length: $\xi=\sqrt{\frac{\hbar^{2}}{2m_{B}N_{0}g_{BB}}%
}$. The interaction amplitude $V_{\vec{k}}$ is given by:%
\begin{equation}
V_{\vec{k}}=\sqrt{N_{0}}g_{IB}\left(  \frac{\left(  \xi k\right)  ^{2}%
}{\left(  \xi k\right)  ^{2}+2}\right)  ^{1/4}.\label{IntAmp}%
\end{equation}

\section{Polaronic ground state properties in $d$ dimensions}

In this section we summarize the main results from standard polaron theory
regarding the ground state properties with emphasis on the dependency on the
dimension (see for example Ref. \cite{2010arXiv1012.4576D} for more details)
and we apply this to the polaronic system consisting of an impurity in a condensate.

\subsection{Jensen-Feynman variational principle}

The most accurate available description of the ground state properties of a
polaron is based on the Jensen-Feynman inequality which states
\cite{PhysRev.97.660, BoekFeynman}:%
\begin{equation}
\mathcal{F}\leq \mathcal{F}_{0}+\frac{1}{\hbar \beta}\left \langle \mathcal{S-S}%
_{0}\right \rangle _{\mathcal{S}_{0}},\label{JensFey}%
\end{equation}
with $\mathcal{F}$ the free energy of the system, $\mathcal{F}_{0}$ the free
energy of a trial system, $\beta=\left(  k_{B}T\right)  ^{-1}$ the inverse
temperature with $k_{B}$ the Boltzmann constant,$\  \mathcal{S}$ the action of
the system and $\mathcal{S}_{0}$ the action of the trial system. It was
suggested by Feynman to consider the particle harmonically coupled to a mass
$M$ with a coupling constant $MW^{2}$ for the trial system
\cite{PhysRev.97.660}. This leads to the following expression for the
Jensen-Feynman inequality (\ref{JensFey}) \cite{PhysRevB.33.3926}:%
\begin{align}
\mathcal{F} &  \leq \frac{d}{\beta}\left \{  \ln \left[  2\sinh \left(
\frac{\beta \hbar \Omega}{2}\right)  \right]  -\ln \left[  2\sinh \left(
\frac{\beta \hbar \Omega}{2\sqrt{1+M/m}}\right)  \right]  \right \}  \nonumber \\
&  -\frac{1}{\beta}\ln \left[  V\left(  \frac{m+M}{2\pi \hbar^{2}\beta}\right)
^{d/2}\right]  -\frac{d}{2\beta}\frac{M}{m+M}\left[  \frac{\hbar \beta \Omega
}{2}\coth \left[  \frac{\hbar \beta \Omega}{2}\right]  -1\right]  \nonumber \\
&  -\sum_{\vec{k}}\frac{\left \vert V_{\vec{k}}\right \vert ^{2}}{\hbar}\int
_{0}^{\hbar \beta/2}du\mathcal{G}\left(  \vec{k},u\right)  \mathcal{M}%
_{M,\Omega}\left(  \vec{k},u\right)  ,\label{JensFeyn}%
\end{align}
with $d$ the dimension, $V$ the volume, $\Omega=W\sqrt{1+M/m_{I}}$ and
$\mathcal{G}\left(  \vec{k},u\right)  $ the Green function of the Bogoliubov
excitations:%
\begin{equation}
\mathcal{G}\left(  \vec{k},u\right)  =\frac{\cosh \left[  \omega_{\vec{k}%
}\left(  u-\hbar \beta/2\right)  \right]  }{\sinh \left[  \hbar \beta \omega
_{\vec{k}}/2\right]  },
\end{equation}
and $\mathcal{M}_{M,\Omega}\left(  \vec{k},u\right)  $ the memory function:%
\begin{equation}
\mathcal{M}_{M,\Omega}\left(  \vec{k},u\right)  =\exp \left[  -\frac{\hbar
k^{2}}{2\left(  m+M\right)  }\left \{  u-\frac{u^{2}}{\hbar \beta}-\frac{M}%
{m}\frac{\cosh \left[  \Omega \hbar \beta/2\right]  -\cosh \left[  \Omega \left(
\hbar \beta/2-u\right)  \right]  }{\Omega \sinh \left(  \hbar \beta \Omega
/2\right)  }\right \}  \right]  .
\end{equation}
The parameters $\Omega$ and $M$ are then determined variationally by
minimizing the expression (\ref{JensFeyn}). The present treatment also allows
an estimation of the radius of the polaron as the root mean square of the
reduced coordinate $\vec{r}$ of the model system \cite{PhysRevB.31.4890}:%
\begin{equation}
\left \langle r^{2}\right \rangle =d\frac{\hbar}{2\Omega}\frac{m_{I}+M}{Mm_{I}%
}\coth \left(  \frac{\beta \hbar \Omega}{2}\right)  .\label{radius}%
\end{equation}
In \cite{PhysRev.97.660} Feynman also presented a calculation of the polaronic
effective mass $m^{\ast}$ at zero temperature:%
\begin{equation}
m^{\ast}=m_{I}+\frac{1}{d}\sum_{\vec{k}}k^{2}\frac{\left \vert V_{\vec{k}%
}\right \vert ^{2}}{\hbar}\int_{0}^{\infty}due^{-\omega_{\vec{k}}u}%
\mathcal{F}_{M,\Omega}\left(  \vec{k},u\right)  u^{2},\label{EffMassFeyT0}%
\end{equation}
with:%
\begin{equation}
\mathcal{F}_{M,\Omega}\left(  \vec{k},u\right)  =\lim_{\beta \rightarrow \infty
}\mathcal{M}_{M,\Omega}\left(  \vec{k},u\right)  =\exp \left \{  -\frac{\hbar
k^{2}}{2\left(  m+M\right)  \Omega}\left[  \frac{M}{m}\left(  1-e^{-\Omega
u}\right)  +\Omega u\right]  \right \}  .
\end{equation}
As far as we know there exists no generalization of equation
(\ref{EffMassFeyT0}) to finite temperatures but as a first estimation we use
(\ref{EffMassFeyT0}) with the temperature dependent variational parameters $M$
and $\Omega$.

\subsection{Polaron consisting of an impurity in a condensate}

Here we introduce the Bogoliubov spectrum (\ref{BogSpectr}) and the
interaction amplitude (\ref{IntAmp}) which are specific for the polaronic
system consisting of an impurity in a condensate. This allows us to write the
Jensen-Feynman inequality (\ref{JensFeyn}) as (we also use polaronic units,
i.e. $\hbar=\xi=m_{I}=1$):%
\begin{align}
\mathcal{F} &  \leq \frac{d}{\beta}\left \{  \ln \left[  2\sinh \left(
\frac{\beta \Omega}{2}\right)  \right]  -\ln \left[  2\sinh \left(  \frac
{\beta \Omega}{2\sqrt{1+M}}\right)  \right]  \right \}  \nonumber \\
&  -\frac{1}{\beta}\ln \left[  V\left(  \frac{1+M}{2\pi \beta}\right)
^{d/2}\right]  -\frac{d}{2\beta}\frac{M}{1+M}\left[  \frac{\beta \Omega}%
{2}\coth \left[  \frac{\beta \Omega}{2}\right]  -1\right]  \nonumber \\
&  -\frac{\alpha^{\left(  d\right)  }}{4\pi}\left(  \frac{m_{B}+1}{m_{B}%
}\right)  ^{2}\int_{0}^{\infty}dk\frac{k^{d}}{\sqrt{k^{2}+2}}\int_{0}%
^{\beta/2}du\mathcal{G}\left(  k,u\right)  \mathcal{M}_{M,\Omega}\left(
k,u\right)  ,\label{JenFeyDim}%
\end{align}
where we introduced the dimensionless coupling parameter $\alpha^{\left(
d\right)  }$ as follows:%
\begin{equation}
\alpha^{\left(  d\right)  }=4\pi \frac{2\pi^{d/2}}{\Gamma \left(  \frac{d}%
{2}\right)  }N_{0}g_{IB}^{2}\left(  \frac{m_{I}\xi^{2}}{\hbar^{2}}\right)
^{2}\frac{V}{\left(  2\pi \xi \right)  ^{d}}\left(  \frac{m_{B}}{m_{B}+m_{I}%
}\right)  ^{2},\label{KopConst}%
\end{equation}
with $\Gamma \left(  x\right)  $ the gamma function. The prefactor was chosen
to be in agreement with the definition for $\alpha^{\left(  3\right)  }$ in
Ref. \cite{PhysRevB.80.184504}\ . Note that the coupling parameter depends on
the impurity-boson interaction amplitude $g_{IB}$ and also on the boson-boson
interaction amplitude $g_{BB}$ through the healing length $\xi$. As mentioned
in the introduction these interaction amplitudes and thus also the coupling
parameter can be externally tuned through a Feshbach resonance or in reduced
dimensions also with a confinement induced resonance.

For $d=2$ the $k$-integral in (\ref{JenFeyDim}) contains an ultraviolet
divergence. This is also the case in 3 dimensions and it was shown in Ref.
\cite{PhysRevB.80.184504} that this is solved by applying the
Lippmann-Schwinger equation up to second order for the interaction amplitude
in the second term of the Hamiltonian (\ref{TotHamilt}). This results in a
renormalization factor that is incorporated through the following substitution
\cite{PhysRevB.80.184504}:%
\begin{equation}
N_{0}g_{IB}\rightarrow N_{0}\left(  T\left(  E\right)  +g_{IB}^{2}\sum
_{\vec{k}}\frac{1}{\frac{\hbar^{2}k^{2}}{2m_{r}}-E}\right)  , \label{Renorm}%
\end{equation}
with $T\left(  E\right)  $ the scattering $T$-matrix. In 2 dimensions the
limit $E\rightarrow0$ in (\ref{Renorm}) results in an infrared divergence. The
second term in (\ref{Renorm}) can be written as:%
\begin{equation}
N_{0}g_{IB}^{2}\sum_{\vec{k}}\frac{1}{\frac{\hbar^{2}k^{2}}{2m_{r}}-E}%
=\frac{\alpha^{\left(  2\right)  }}{2\pi}\frac{\hbar^{2}}{m_{I}\xi^{2}}%
\frac{m_{B}+m_{I}}{m_{B}}\int_{0}^{\infty}\frac{k}{k^{2}-2m_{r}E/\hbar^{2}}dk,
\label{Regul}%
\end{equation}
which lifts the ultraviolet divergence in (\ref{JenFeyDim}). For numerical
considerations a cutoff $K_{c}$ is introduced for the $k$-integral which
enables us to calculate the integral in (\ref{Regul}):%
\begin{align}
\frac{\alpha^{\left(  2\right)  }}{2\pi}\frac{\hbar^{2}}{m_{I}\xi^{2}}%
\frac{m_{B}+m_{I}}{m_{B}}\int_{0}^{K_{c}}\frac{k}{k^{2}-2m_{r}\xi^{2}%
E/\hbar^{2}}dk  &  =\frac{\alpha^{\left(  2\right)  }}{4\pi}\frac{\hbar^{2}%
}{m_{I}\xi^{2}}\frac{m_{B}+m_{I}}{m_{B}}\ln \left(  \frac{\frac{\hbar^{2}%
K_{c}^{2}}{2m_{r}}-E}{E}\right) \nonumber \\
&  \approx \frac{\alpha^{\left(  2\right)  }}{4\pi}\frac{\hbar^{2}}{m_{I}%
\xi^{2}}\frac{m_{B}+m_{I}}{m_{B}}\ln \left(  \frac{\hbar^{2}K_{c}^{2}}{2m_{r}%
E}\right)  , \label{Regular}%
\end{align}
where in the second step we used the fact that the energy related to the
cutoff is much larger than the typical energy of the scattering event $E$.
Equation (\ref{Regular}) shows that the chosen value of $E$ is not important
since it only results in an energy shift and therefore has no influence on the
physical properties of the system.

\subsection{Results}

We apply the presented treatment to the system of a lithium-6 impurity in a
sodium condensate ($m_{B}/m_{I}=3.82207$). All results are presented in
polaronic units, i.e. $\hbar=\xi=m_{I}=1$.

\begin{figure}
[ptb]
\begin{center}
\includegraphics[
height=8.5361cm,
width=8.7052cm
]%
{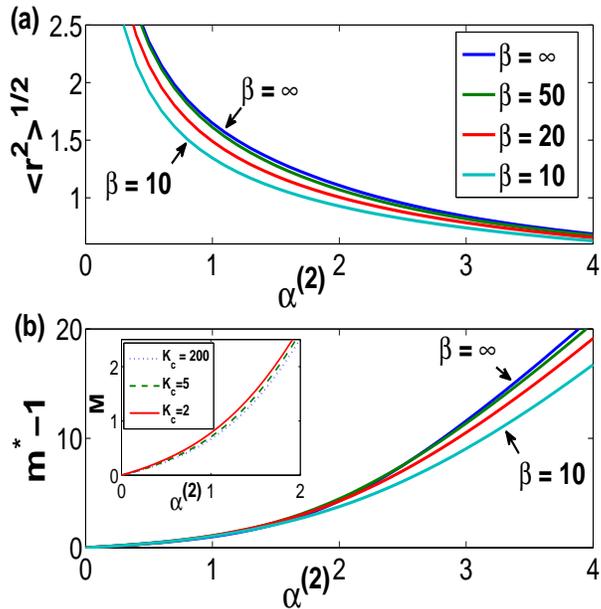}%
\caption{(Color online) The groundstate properties of the polaron consisting
of a lithium-6 impurity in a sodium condensate in 2 dimensions. In (a) the
radius of the polaron (\ref{radius}) is presented and in (b) the effective
mass (\ref{EffMassFeyT0}) as a function of the polaronic coupling parameter
$\alpha^{\left(  2\right)  }$ at different temperatures ($\beta=\left(
k_{B}T\right)  ^{-1}$) and with a cutoff $K_{c}=200$. The inset shows the
variational parameter $M$ for different values of the cutoff $K_{c}$ at
$\beta=50$. All results are presented in polaronic units ($\hbar=m_{I}=\xi
=1$). }%
\label{fig: Grond2D}%
\end{center}
\end{figure}

In figure \ref{fig: Grond2D} the results for the polaronic ground state
properties in 2 dimensions as a function of the coupling parameter
$\alpha^{\left(  2\right)  }$ are presented. In (a) the radius of the polaron
is shown and in (b) the effective mass at different temperatures. The observed
behavior is analogous to the three-dimensional case (see Ref.
\cite{PhysRevB.80.184504})\ and suggests that for growing $\alpha^{\left(
2\right)  }$ the self-induced potential becomes stronger leading to a bound
state at high enough $\alpha^{\left(  2\right)  }$. However, as compared to
the three-dimensional case, the transition is much smoother with a transition
region between $\alpha^{\left(  2\right)  }\approx1$ and $\alpha^{\left(
2\right)  }\approx3$. This behavior is in agreement with the mean-field
results of Refs. \cite{PhysRevB.46.301, 0295-5075-82-3-30004}, where also a
smooth transition to the self-trapped state was found for $d=2$. For the
cutoff $K_{c}$ we used the inverse of the Van der Waals radius for sodium
which results in $K_{c}=200$. To check whether this cutoff is large enough the
variational parameter $M$ is plotted in the inset of figure
(\ref{fig: Grond2D})(b) for different values of $K_{c}$ which reveals already
a reasonable convergence at $K_{c}\approx5$.%

\begin{figure}
[ptb]
\begin{center}
\includegraphics[
height=8.5361cm,
width=8.703cm
]%
{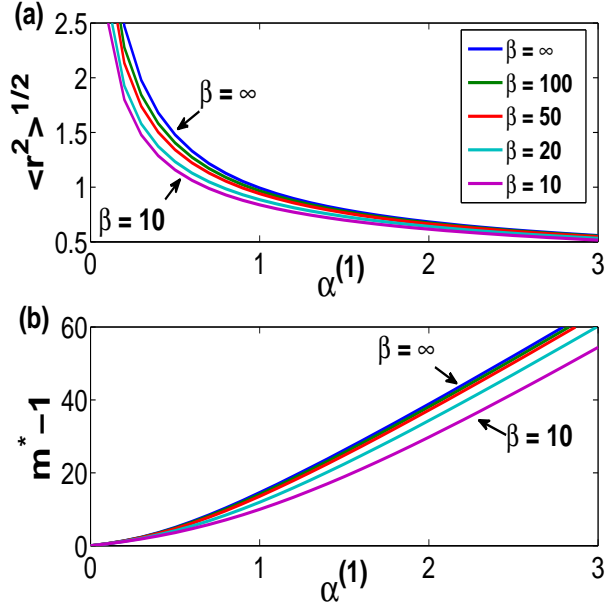}%
\caption{(Color online) The polaronic ground state properties of a lithium-6
impurity in a sodium condensate in 1 dimension. The radius (a) and the
effective mass (b) are presented as a function of the polaronic coupling
parameter $\alpha^{\left(  1\right)  }$ at different temperatures
($\beta=\left(  k_{B}T\right)  ^{-1}$). All results are presented in polaronic
units ($\hbar=m_{I}=\xi=1$).}%
\label{fig: grond1D}%
\end{center}
\end{figure}

In figure \ref{fig: grond1D} the results for the 1-dimensional case are
presented. In (a) the radius of the polaron is plotted and (b) shows the
effective mass at different temperatures. For growing $\alpha^{\left(
1\right)  }$ the characteristics of the appearance of a bound state in the
self-induced potential are again observed. The characteristics of the weak
coupling regime are however not present and the transition region is between
$\alpha^{\left(  1\right)  }=0$ and $\alpha^{\left(  1\right)  }\approx1$.
This is again in agreement with the mean-field results of Refs.
\cite{PhysRevB.46.301, 0295-5075-82-3-30004} for $d=1$.

\section{Response to Bragg scattering in $d$ dimensions}

The response of a system to Bragg spectroscopy is proportional to the
imaginary part of the density response function (\ref{DensResp}). In Ref.
\cite{PhysRevA.83.033631} it was shown that the use of the
Feynman-Hellwarth-Iddings-Platzman approximation (as introduced in Ref.
\cite{PhysRev.127.1004}\ for a calculation of the impedance of the
Fr\"{o}hlich solid state polaron and generalized in Ref.
\cite{PhysRevB.5.2367} for the optical absorption) leads to the following
expression for the Bragg response:
\begin{equation}
\operatorname{Im}\left[  \chi \left(  \omega,\vec{k}\right)  \right]
=-\frac{k^{2}}{m_{I}}\frac{\operatorname{Im}\left[  \Sigma \left(  \omega
,\vec{k}\right)  \right]  }{\left \{  \omega^{2}-\operatorname{Re}\left[
\Sigma \left(  \omega,\vec{k}\right)  \right]  \right \}  ^{2}+\left \{
\operatorname{Im}\left[  \Sigma \left(  \omega,\vec{k}\right)  \right]
\right \}  ^{2}}, \label{BraggResp}%
\end{equation}
with $\Sigma \left(  \omega,\vec{k}\right)  $ the self energy:%
\begin{align}
\Sigma \left(  \omega,\vec{k}\right)   &  =\frac{2}{m_{I}N\hbar}\sum_{\vec
{q}\neq0}\left \vert V_{\vec{q}}\right \vert ^{2}\frac{\left(  \vec{k}.\vec
{q}\right)  ^{2}}{k^{2}}\label{SelfEnergy}\\
&  \times \int_{0}^{\infty}dt\left(  1-e^{i\omega t}\right)  \operatorname{Im}%
\left \{  \left[  e^{i\omega_{\vec{q}}t}+2\cos \left(  \omega_{\vec{q}}t\right)
n\left(  \omega_{\vec{q}}\right)  \right]  \exp \left[  -\left(  \vec{k}%
+\vec{q}\right)  ^{2}D\left(  t\right)  \right]  \right \}  ,
\end{align}
$n\left(  \omega \right)  =\left(  \exp \left[  \beta \hbar w\right]  -1\right)
^{-1}$ the Bose-Einstein distribution and:%
\begin{equation}
D\left(  t\right)  =\frac{t^{2}}{2\beta \left(  m_{I}+M\right)  }-i\frac{\hbar
}{2\left(  m_{I}+M\right)  }t+\frac{\hbar M}{2m_{I}\Omega \left(
m_{I}+M\right)  }\left[  1-e^{i\Omega t}+4\sin^{2}\left(  \frac{\Omega t}%
{2}\right)  n\left(  \Omega \right)  \right]  . \label{Dfun}%
\end{equation}
For numerical calculations the representation for the self energy as derived
in appendix \ref{App: AndereRep} is used.

\subsection{Sum rule}

As was first noted in \cite{PhysRevB.15.1212} for the Fr\"{o}hlich polaron and
generalized in \cite{PhysRevA.83.033631} for an impurity in a condensate the
f-sum rule can be written as:%
\begin{equation}
\frac{\pi}{2}\frac{1}{\left(  1-R\left(  \alpha,k\right)  \right)  }%
+\frac{m_{I}}{k^{2}}\int_{\varepsilon}^{\infty}d\omega \omega \operatorname{Im}%
\left[  \chi \left(  \omega,\vec{k}\right)  \right]  =\frac{\pi}{2},
\label{Somregel}%
\end{equation}
with $\varepsilon$ a small number such that the Drude peak (see later) is not
included in the integral and:%
\begin{equation}
R\left(  \alpha,k\right)  =\lim_{\omega \rightarrow0}\frac{\operatorname{Re}%
\left[  \Sigma \left(  \omega,\vec{k}\right)  \right]  }{\omega^{2}}.
\label{RFun}%
\end{equation}
In the limits $\beta \rightarrow \infty$ and $k\rightarrow0$ the function
(\ref{RFun}) is related to the Feynman effective mass (\ref{EffMassFeyT0})
\cite{PhysRevB.15.1212}:%
\begin{equation}
m^{\ast}=m_{I}\left(  1-\lim_{\beta \rightarrow \infty}R\left(  \alpha,0\right)
\right)  .
\end{equation}
This relation provides a powerful experimental tool to determine the effective
mass from the optical response which was recently applied for the Fr\"{o}hlich
solid state polaron \cite{PhysRevB.81.125119, PhysRevLett.100.226403}.

\subsection{Self energy for an impurity in a condensate}

Introducing the interaction amplitude (\ref{IntAmp}) and the coupling
parameters (\ref{KopConst}) in expressions (\ref{ImagZelf}) and (\ref{ReZelf})
for the imaginary and real part of the self energy results in (using polaronic
units):%
\begin{align}
\operatorname{Im}\left[  \Sigma \left(  \omega,\vec{k}\right)  \right]   &
=\sqrt{2\pi \beta \left(  1+M\right)  }\frac{\alpha^{\left(  d\right)  }}{8\pi
}\frac{\Gamma \left(  d/2\right)  }{2\pi^{d/2}}\left(  \frac{m_{B}+1}{m_{B}%
}\right)  ^{2}B\left(  \beta,n,n^{\prime}\right) \nonumber \\
&  \times \sum_{n,n^{\prime}=0}^{\infty}\int d\vec{q}\frac{q}{\sqrt{q^{2}+2}%
}\frac{\left(  \vec{k}.\vec{q}\right)  ^{2}}{k^{2}}\left \vert \vec{k}+\vec
{q}\right \vert ^{2\left(  n+n^{\prime}\right)  -1}e^{-a^{2}\left(
\beta \right)  \left(  \vec{k}+\vec{q}\right)  ^{2}}\nonumber \\
&  \times \left \{  \left[  1+n\left(  \omega_{\vec{q}}\right)  \right]  \left[
e^{-\frac{\beta \left(  1+M\right)  \left(  A^{+}+\omega \right)  ^{2}}{2\left(
\vec{k}+\vec{q}\right)  ^{2}}}-e^{-\frac{\beta \left(  1+M\right)  \left(
A^{+}-\omega \right)  ^{2}}{2\left(  \vec{k}^{\prime}+\vec{q}\right)  ^{2}}%
}\right]  \right. \nonumber \\
&  \left.  +n\left(  \omega_{\vec{q}}\right)  \left[  e^{-\frac{\beta \left(
1+M\right)  \left(  A^{-}+\omega \right)  ^{2}}{2\left(  \vec{k}+\vec
{q}\right)  ^{2}}}-e^{-\frac{\beta \left(  1+M\right)  \left(  A^{-}%
-\omega \right)  ^{2}}{2\left(  \vec{k}+\vec{q}\right)  ^{2}}}\right]
\right \}  ;\label{ImagSelf}\\
\operatorname{Re}\left[  \Sigma \left(  \omega,\vec{k}\right)  \right]   &
=\sqrt{2\beta \left(  1+M\right)  }\frac{\alpha^{\left(  d\right)  }}{4\pi
}\frac{\Gamma \left(  d/2\right)  }{2\pi^{d/2}}\left(  \frac{m_{B}+1}{m_{B}%
}\right)  ^{2}B\left(  \beta,n,n^{\prime}\right) \nonumber \\
&  \times \sum_{n,n^{\prime}=0}^{\infty}\int d\vec{q}\frac{q}{\sqrt{q^{2}+2}%
}\frac{\left(  \vec{k}.\vec{q}\right)  ^{2}}{k^{2}}\left \vert \vec{k}+\vec
{q}\right \vert ^{2\left(  n+n^{\prime}\right)  -1}e^{-a^{2}\left(
\beta \right)  \left(  \vec{k}+\vec{q}\right)  ^{2}}\nonumber \\
&  \times \left \{  \left[  1+n\left(  \omega_{\vec{q}}\right)  \right]  \left[
2F\left(  \sqrt{\frac{\beta \left(  m+M\right)  }{2}}\frac{A^{+}}{\left \vert
\vec{k}+\vec{q}\right \vert }\right)  -F\left(  \sqrt{\frac{\beta \left(
m+M\right)  }{2}}\frac{A^{+}+\omega}{\left \vert \vec{k}+\vec{q}\right \vert
}\right)  \right.  \right. \nonumber \\
&  \left.  -F\left(  \sqrt{\frac{\beta \left(  m+M\right)  }{2}}\frac
{A^{+}-\omega}{\left \vert \vec{k}+\vec{q}\right \vert }\right)  \right]
+n\left(  \omega_{\vec{q}}\right)  \left[  2F\left(  \sqrt{\frac{\beta \left(
m+M\right)  }{2}}\frac{A^{-}}{\left \vert \vec{k}+\vec{q}\right \vert }\right)
\right. \nonumber \\
&  \left.  \left.  -F\left(  \sqrt{\frac{\beta \left(  m+M\right)  }{2}}%
\frac{A^{-}+\omega}{\left \vert \vec{k}+\vec{q}\right \vert }\right)  -F\left(
\sqrt{\frac{\beta \left(  m+M\right)  }{2}}\frac{A^{-}-\omega}{\left \vert
\vec{k}+\vec{q}\right \vert }\right)  \right]  \right \}  .
\end{align}
See appendix \ref{App: AndereRep} for the definition of the different
functions. These expressions are suited for numerical calculations of the
Bragg response.

\subsection{Weak coupling limit}

At weak polaronic coupling the Bragg response (\ref{BraggResp}) to lowest
order in $\alpha$ is given by (in polaronic units):%
\begin{equation}
\operatorname{Im}\left[  \chi^{W}\left(  \omega,\vec{k}\right)  \right]
=-\frac{k^{2}}{\omega^{4}}\operatorname{Im}\left[  \Sigma^{W}\left(
\omega,\vec{k}\right)  \right]  .
\end{equation}
In the weak coupling limit the variational parameter $M$ tends to zero and the
imaginary part of the self energy (\ref{ImagSelf}) becomes:%
\begin{align}
\operatorname{Im}\left[  \Sigma^{W}\left(  \omega,\vec{k}\right)  \right]   &
=\frac{\sqrt{2\beta \pi}}{2}\sum_{\vec{q}\neq0}\left \vert V_{\vec{q}%
}\right \vert ^{2}\frac{\left(  \vec{k}.\vec{q}\right)  ^{2}}{k^{2}}\left \vert
\vec{k}+\vec{q}\right \vert ^{-1}\nonumber \\
&  \times \left(
\begin{array}
[c]{c}%
\left[  1+n\left(  \omega_{\vec{q}}\right)  \right]  \left \{  \exp \left[
-\frac{2\beta \left(  B^{+}+\omega \right)  ^{2}}{4\left(  \vec{k}+\vec
{q}\right)  ^{2}}\right]  -\exp \left[  -\frac{2\beta \left(  B^{+}%
-\omega \right)  ^{2}}{4\left(  \vec{k}+\vec{q}\right)  ^{2}}\right]  \right \}
\\
+n\left(  \omega_{\vec{q}}\right)  \left \{  \exp \left[  -\frac{2\beta \left(
B^{-}+\omega \right)  ^{2}}{4\left(  \vec{k}+\vec{q}\right)  ^{2}}\right]
-\exp \left[  -\frac{2\beta \left(  B^{-}-\omega \right)  ^{2}}{4\left(  \vec
{k}+\vec{q}\right)  ^{2}}\right]  \right \}
\end{array}
\right)  , \label{ImagZwak}%
\end{align}
with:%
\begin{equation}
B^{\pm}=\pm \omega_{\vec{q}}+\frac{\left(  \vec{k}+\vec{q}\right)  ^{2}}{2}.
\label{Aplusmin}%
\end{equation}
These expressions coincide with the weak coupling result obtained in the
framework of Gurevich, Lang and Firsov \cite{Gurevich}.

\subsection{Results}

We present the Bragg response for a lithium-6 impurity in a sodium condensate
($m_{B}/m_{I}=3.82207$). All results are presented in polaronic units, i.e.
$\hbar=\xi=m_{I}=1$.

\begin{figure}
[ptb]
\begin{center}
\includegraphics[
height=2.527in,
width=5.0721in
]%
{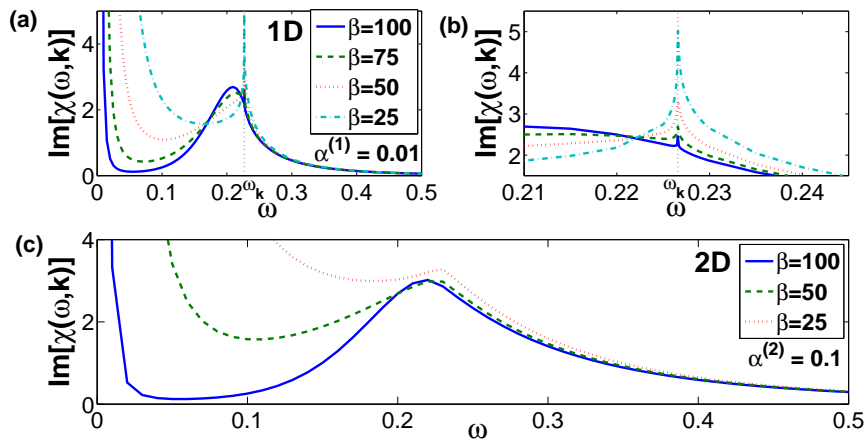}%
\caption{(Color online) The Bragg response (\ref{BraggResp}) at weak polaronic
coupling, momentum exchange $k=1$ and for different temperatures
($\beta=\left(  k_{B}T\right)  ^{-1}$) in 1D (a) and 2D (c). In both cases a
peak corresponding to the emission of Bogoliubov excitations is observed
together with the anomalous Drude peak at $\omega=0$. In 1D another sharp peak
is present at $\omega=\omega_{k}$, with $\omega_{k}$ the Bogoliubov dispersion
(\ref{BogSpectr}). In (b) we have zoomed in on this sharp peak in 1D. All
quantities are in polaronic units ($\hbar=m_{I}=\xi=1$).}%
\label{fig: RespTafh}%
\end{center}
\end{figure}

In figure \ref{fig: RespTafh} the Bragg response (\ref{BraggResp}) is
presented for different temperatures and for a momentum exchange $k=1$ in 1
and 2 dimensions at weak polaronic coupling ($\alpha^{\left(  1\right)
}=0.01$ and $\alpha^{\left(  2\right)  }=0.1$). In both cases we observe the
Drude peak centered at $\omega=0$ and a peak corresponding to the emission of
Bogoliubov excitations. This is qualitatively the same behavior as in the
three-dimensional case \cite{PhysRevA.83.033631}, quantitatively we observe
that the amplitude of the Bogoliubov emission peak increases as the dimension
is reduced. The Drude peak is a well-known feature in the response spectra of
the Fr\"{o}hlich polaron (see for example Refs. \cite{Huybrechts1973163,
PhysRevLett.100.226403, springerlink:10.1007/PL00011092}) and is a consequence
of the incoherent scattering of the polaron with thermal Bogoliubov
excitations. The width of the Drude peak scales with the scattering rate for
absorption of a Bogoliubov excitation which is proportional to the number of
thermally excited Bogoliubov excitations \cite{Mahan}. This explains the
temperature dependence of the width of the Drude peak in figure
\ref{fig: RespTafh}.

In 1D another sharp peak is observed in figure \ref{fig: RespTafh} at
$\omega=\omega_{k}$ (with $\omega_{k}$ the Bogoliubov dispersion
(\ref{BogSpectr}) and $k$ the exchanged momentum) which broadens as the
temperature is increased and dominates the Bogoliubov emission peak at
relatively high temperatures. This extra peak in 1D is associated with the
weak coupling regime since at intermediate coupling the sharp structure
disappears and the peak merges with the Bogoliubov emission peak. The location
indicates that it corresponds to the process where both the exchanged energy
$\hbar \omega$ and momentum $\vec{k}$ are transferred to a Bogoliubov
excitation. Whether this extra peak is experimentally observable is
questionable since it is only visible at relatively high temperatures where in
reduced dimensions thermal phase fluctuations can become important and destroy
the polaronic features.

\begin{figure}
[ptb]
\begin{center}
\includegraphics[
height=6.2077cm,
width=12.8678cm
]%
{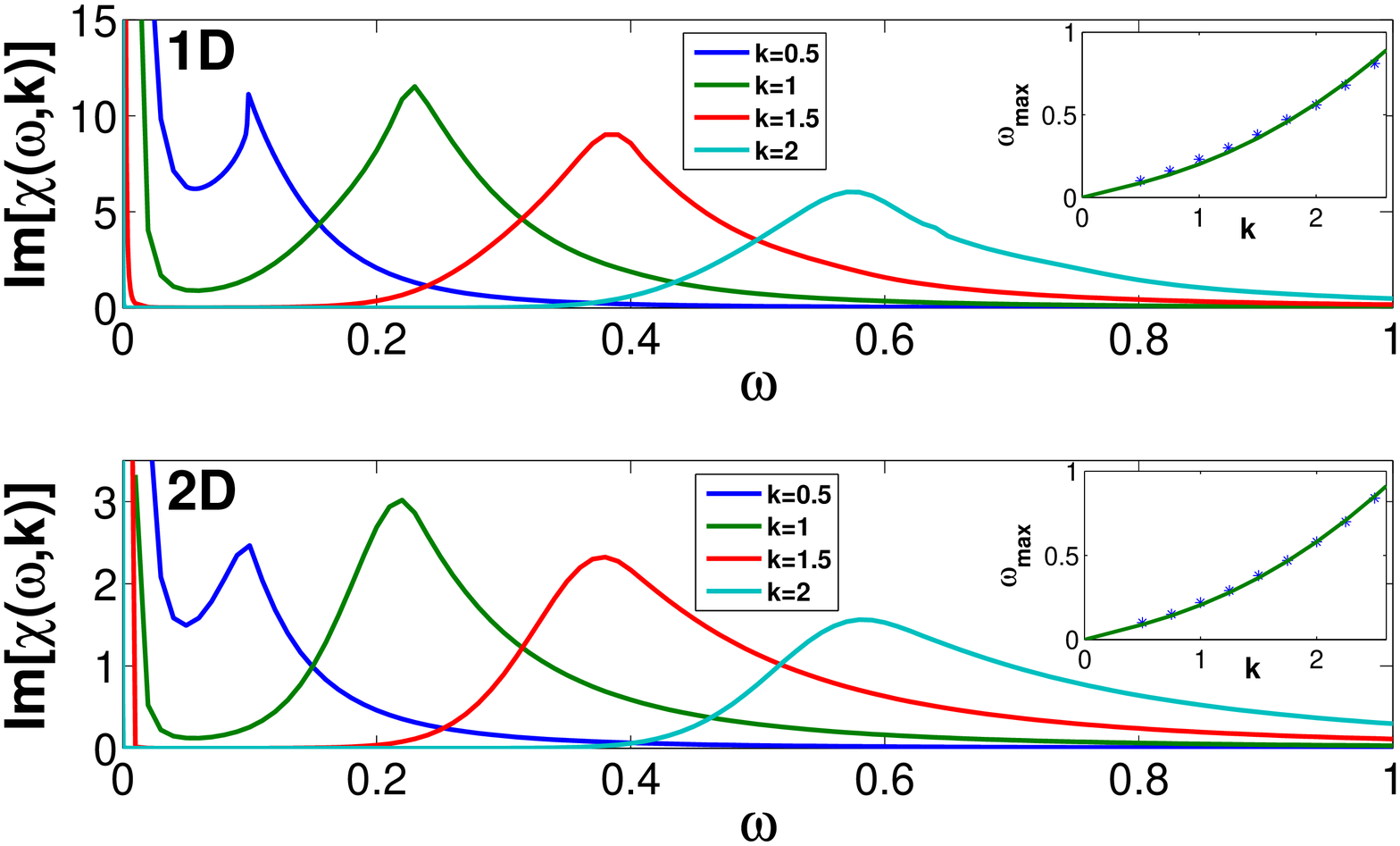}%
\caption{(Color online) The Bragg response at weak polaronic coupling
($\alpha^{\left(  d\right)  }=0.1$) for different exchanged momenta $k$ in 1
and 2 dimensions and at a temperature $\beta=100$. The inset shows the
location of the maximum of the peak as a function of the exchanged momentum
(markers) together with a least square fit to the Bogoliubov spectrum
(\ref{BogSpectr}) (full line), this results in $m_{B}=4.3159$ ($4.2216$) for
the fitting parameter in 1D (2D). Everything is in polaronic units
($\hbar=m_{I}=\xi=1$).}%
\label{Fig: ZwakKAfh}%
\end{center}
\end{figure}

Figure \ref{Fig: ZwakKAfh} presents the Bragg response for different momenta
exchange at a temperature $\beta=100$ (where the sharp peak at the Bogoliubov
dispersion in 1D is too narrow to perceive). The insets show the location of
the maximum of the Bogoliubov emission peak as a function of the exchanged
momentum together with a least square fit to the Bogoliubov spectrum
(\ref{BogSpectr}) which results in a good agreement. The optimal fitting
parameter is determined as $m_{B}=4.3159$ ($4.2216$) in 1D (2D).%

\begin{figure}
[ptb]
\begin{center}
\includegraphics[
height=5.9485cm,
width=11.6333cm
]%
{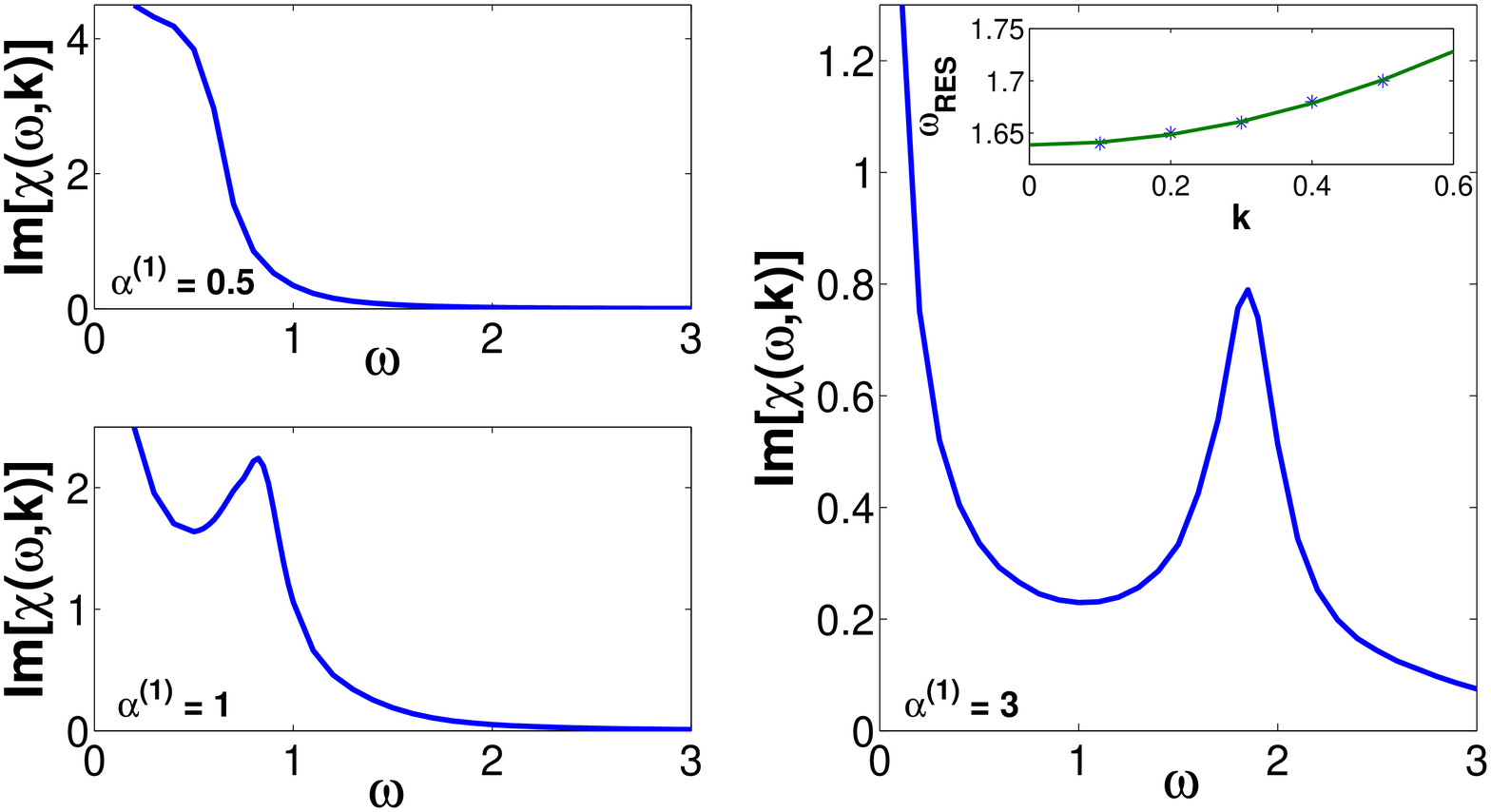}%
\caption{(Color online) Here we zoomed in on the tail of the Bogoliubov
emission peak for momentum exchange $k=1$ and temperature $\beta=100$ in 1
dimension. It is clear that at larger values for $\alpha^{\left(  1\right)  }$
a secondary peak emerges. The inset shows the location of the maximum of this
secondary peak at $\alpha^{\left(  1\right)  }=3$ as a function of the
exchanged momentum (markers) together with a least square fit to a quadratic
spectrum (\ref{KwadSpectr}) (solid line), this results in $\omega=1.6386$ and
$m=2.0107$ for the fitting parameters. Everything is in polaronic units
($\hbar=m_{I}=\xi=1$).}%
\label{Fig: RES1D}%
\end{center}
\end{figure}

\begin{figure}
[ptb]
\begin{center}
\includegraphics[
height=5.7705cm,
width=11.5279cm
]%
{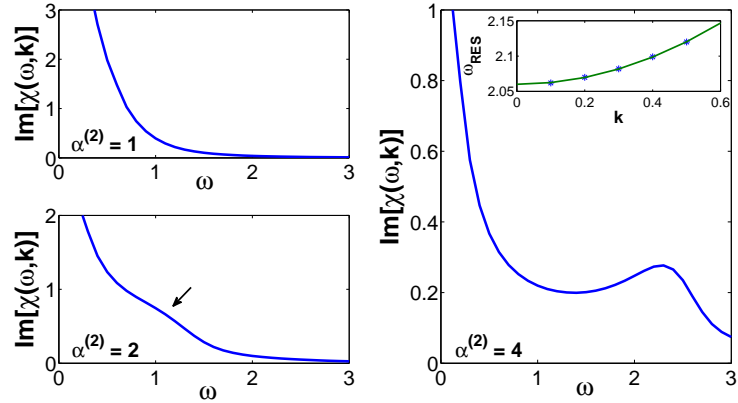}%
\caption{(Color online) Here we zoomed in on the tail of the Bogoliubov
emission peak for momentum exchange $k=1$ and temperature $\beta=100$ in 2
dimensions. As in the one-dimensional case a secondary peak emerges at larger
values for $\alpha^{\left(  2\right)  }$. The inset shows the location of the
maximum of this secondary peak at $\alpha^{\left(  2\right)  }=4$ as a
function of the exchanged momentum (markers) together with a least square fit
to a quadratic spectrum (\ref{KwadSpectr}) (solid line), this results in
$\omega=2.0601$ and $m=2.0755$ for the fitting parameters. Everything is in
polaronic units ($\hbar=m_{I}=\xi=1$).}%
\label{Fig: RES2D}%
\end{center}
\end{figure}

In figures \ref{Fig: RES1D} and \ref{Fig: RES2D} we have zoomed in on the tail
of the Bogoliubov emission peak for different values of the coupling parameter
in 1 and 2 dimensions, respectively. At larger values for the polaronic
coupling parameter $\alpha^{\left(  d\right)  }$ the emergence of a secondary
peak is observed. This behavior is also observed in the optical absorption of
the Fr\"{o}hlich solid state polaron where the secondary peak corresponds to a
transition to the Relaxed Excited State accompanied by the emission of phonons
\cite{PhysRevLett.22.94}. The Relaxed Excited State denotes an excitation of
the impurity in the relaxed self-induced potential where relaxed means that
the self-induced potential is adapted to the excited state wave function of
the impurity. In the inset the location of this secondary peak is plotted as a
function of the exchanged momentum together with a least square fit to the
following quadratically spectrum:
\begin{equation}
\omega \left(  k\right)  =\omega+\frac{\hbar^{2}k^{2}}{2m},\label{KwadSpectr}%
\end{equation}
which shows a good agreement at small $k$. This suggests that the state
corresponding to the secondary peak is characterized by a transition frequency
$\omega$ and an effective mass $m$ (this was also observed for the
3-dimensional case in Ref. \cite{PhysRevA.83.033631}).%

Finally we have checked whether the spectra satisfy the sum rule
(\ref{Somregel}). We calculated the sum of the two terms on the left hand side
of expression (\ref{Somregel}) which is presented in table
\ref{Tab: Somregel1d} for $d=1$ and in table \ref{Tab: Somregel2d} for $d=2$
at $\beta=100$ and at different values for $\alpha$ and $k$. These values
should be compared to $\pi/2=1.5708$ which results in a fair agreement with
small deviations which are to be expected since numerically we had to
introduce a cutoff for the $\omega$-integral in (\ref{Somregel}) and the
choice of the parameter $\varepsilon$ in (\ref{Somregel}) is somewhat
arbitrary resulting in a double counting of part of the weight of the Drude peak.%

\begin{table}[tbp] \centering
$%
\begin{tabular}
[c]{l||ll}
& $\alpha^{\left(  1\right)  }=0.1$ & $\alpha^{\left(  1\right)  }%
=3$\\ \hline \hline
$k=1$ & $1.5440$ & $1.5547$\\
$k=3$ & $1.5544$ & $1.5743$%
\end{tabular}
\  \  \ $%
\caption{Here we show the sum of the two terms at the left hand side of the f-sum rule (\ref{Somregel}) in 1 dimension at $\beta = 100$ and at different values for $\alpha^{(1)}$ and $k$.}\label{Tab: Somregel1d}%
\end{table}%
%

\begin{table}[tbp] \centering
$%
\begin{tabular}
[c]{l||ll}
& $\alpha^{\left(  2\right)  }=1$ & $\alpha^{\left(  2\right)  }%
=4$\\ \hline \hline
$k=1$ & $1.5678$ & $1.5734$\\
$k=3$ & $1.5669$ & $1.5800$%
\end{tabular}
\  \  \  \ $%
\caption{Here we show the sum of the two terms at the left hand side of the f-sum rule (\ref{Somregel}) in 2 dimensions at $\beta = 100$ and at different values for $\alpha^{(2)}$ and $k$.}\label{Tab: Somregel2d}%
\end{table}%

\section{Discussion and conclusions}

We have applied the calculations for the polaronic ground state properties of
an impurity in a Bose-Einstein condensate and the response of this system to
Bragg spectroscopy to reduced dimensions. For this purpose we introduced a
polaronic coupling parameter $\alpha^{\left(  d\right)  }$ (\ref{KopConst})
which depends on the dimension. For growing $\alpha^{\left(  d\right)  }$ the
ground state properties suggest that the self-induced potential accomodates a
bound state. As compared to the three-dimensional case the transition to the
self-trapped state is much smoother in reduced dimension and for $d=1$ the
characteristics of the weak coupling regime are absent.

The Bragg response of the system revealed a peak corresponding to the emission
of Bogoliubov excitations, the Drude peak and the emergence of a secondary
peak in the strong coupling regime. The amplitude of these polaronic features
grows when we go to reduced dimensions. This is important since this indicates
that going to reduced dimensions can facilitate an experimental detection of
polaronic features. In 1D another sharp peak is observed at weak polaronic
coupling that corresponds to the full transition of the exchanged energy and
momentum to a Bogoliubov excitation.

Another advantage of considering reduced dimensions is the possibility of
using confinement-induced resonances which permits a tuning of the polaronic
coupling parameter. These results show that considering an impurity in a
Bose-Einstein condensate in reduced dimensions is a very promising candidate
to experimentally probe the polaronic strong coupling regime for the first time.

\begin{acknowledgments}
The authors gratefully acknowledge fruitful discussions with M. Wouters and A.
Widera. This work was supported by FWO-V under Projects No. G.0180.09N, No.
G.0115.06, No. G.0356.06, No. G.0370.09N and G.0119.12N, and the WOG Belgium
under Project No. WO.033.09N. J.T. gratefully acknowledges support of the
Special Research Fund of the University of Antwerp, BOF NOI UA 2004. W.C.
acknowledges financial support from the BOF-UA.
\end{acknowledgments}

\appendix{}

\section{Other representation for the self energy\label{App: AndereRep}}

Here we rewrite the self energy (\ref{SelfEnergy}) to a form which is more
suited for numerical calculations. The presented derivation is based on the
approach for the optical absorption of the Fr\"{o}hlich solid state polaron as
proposed in Refs. \cite{PhysRevB.5.2367, PhysRevB.28.6051}. We start by
rewriting $D\left(  t\right)  $ (\ref{Dfun}) as:
\begin{align}
D\left(  t\right)   &  =\frac{t^{2}}{2\beta \left(  m+M\right)  }-i\frac{\hbar
}{2\left(  m+M\right)  }t\nonumber \\
&  +\frac{\hbar M}{2m\Omega \left(  m+M\right)  }\left \{  \coth \left(
\frac{\hbar \beta \Omega}{2}\right)  -\left[  1+n\left(  \Omega \right)  \right]
e^{i\Omega t}-n\left(  \Omega \right)  e^{-i\Omega t}\right \}  ,
\end{align}
which allows us to write:%
\begin{equation}
e^{-k^{2}D\left(  t\right)  }=e^{-a^{2}\left(  \beta \right)  k^{2}}%
\sum_{n,n^{\prime}}k^{2\left(  n+n^{\prime}\right)  }B\left(  \beta
,n,n^{\prime}\right)  e^{-\frac{k^{2}t^{2}}{2\beta \left(  m+M\right)
}+it\left[  \frac{k^{2}\hbar}{2\left(  m+M\right)  }+\Omega \left(
n-n^{\prime}\right)  \right]  }, \label{DFunAltern}%
\end{equation}
with:%
\begin{align}
a^{2}\left(  \beta \right)   &  =\frac{\hbar M}{2m\Omega \left(  m+M\right)
}\coth \left(  \frac{\hbar \beta \Omega}{2}\right)  ;\nonumber \\
B\left(  \beta,n,n^{\prime}\right)   &  =\frac{1}{n!}\frac{1}{n^{\prime}%
!}\left \{  a^{2}\left[  1+n\left(  \Omega \right)  \right]  \right \}
^{n}\left[  a^{2}n\left(  \Omega \right)  \right]  ^{n^{\prime}};
\end{align}
and $a=a\left(  \infty \right)  $. If we now use (\ref{DFunAltern}) in the
expression for the self energy (\ref{SelfEnergy}) we get:%
\begin{align}
\Sigma \left(  \omega,\vec{k}\right)   &  =\frac{2}{m_{I}N\hbar}\sum
_{n,n^{\prime}=0}^{\infty}\sum_{\vec{q}\neq0}\left \vert V_{\vec{q}}\right \vert
^{2}\frac{\left(  \vec{k}.\vec{q}\right)  ^{2}}{k^{2}}\left \vert \vec{k}%
+\vec{q}\right \vert ^{2\left(  n+n^{\prime}\right)  }B\left(  \beta
,n,n^{\prime}\right)  e^{-a^{2}\left(  \beta \right)  \left(  \vec{k}+\vec
{q}\right)  ^{2}}\nonumber \\
&  \times \int_{0}^{\infty}dt\left(  1-e^{i\omega t}\right)  \operatorname{Im}%
\left \{  \left[  1+n\left(  \omega_{\vec{q}}\right)  \right]  e^{-\frac
{\left(  \vec{k}+\vec{q}\right)  ^{2}}{2\beta \left(  m+M\right)  }t^{2}%
+iA^{+}t}+n\left(  \omega_{\vec{q}}\right)  e^{-\frac{\left(  \vec{k}+\vec
{q}\right)  ^{2}}{2\beta \left(  m+M\right)  }t^{2}+iA^{-}t}\right \}  ,
\label{Zelf2}%
\end{align}
with:%
\begin{equation}
A^{\pm}=\pm \omega_{\vec{q}}+\frac{\left(  \vec{k}+\vec{q}\right)  ^{2}\hbar
}{2\left(  m+M\right)  }+\Omega \left(  n-n^{\prime}\right)  .
\end{equation}
We now split the self energy in an imaginary and a real part. Taking the
imaginary part of (\ref{Zelf2}) results in:%
\begin{align}
\operatorname{Im}\left[  \Sigma \left(  \omega,\vec{k}\right)  \right]   &
=-\frac{2}{m_{I}N\hbar}\sum_{n,n^{\prime}=0}^{\infty}\sum_{\vec{q}\neq
0}\left \vert V_{\vec{q}}\right \vert ^{2}\frac{\left(  \vec{k}.\vec{q}\right)
^{2}}{k^{2}}\left(  \vec{k}+\vec{q}\right)  ^{2\left(  n+n^{\prime}\right)
}B\left(  \beta,n,n^{\prime}\right)  e^{-a^{2}\left(  \beta \right)  \left(
\vec{k}+\vec{q}\right)  ^{2}}\nonumber \\
&  \times \operatorname{Im}\left(  \int_{0}^{\infty}dt\sin \left(  \omega
t\right)  \left \{  \left[  1+n\left(  \omega_{\vec{q}}\right)  \right]
e^{-\frac{\left(  \vec{k}+\vec{q}\right)  ^{2}}{2\beta \left(  m+M\right)
}t^{2}+iA^{+}t}\right.  \right. \nonumber \\
&  \left.  \left.  +n\left(  \omega_{\vec{q}}\right)  e^{-\frac{\left(
\vec{k}+\vec{q}\right)  ^{2}}{2\beta \left(  m+M\right)  }t^{2}+iA^{-}%
t}\right \}  \right)  .
\end{align}
The time integration can now be done (with $C^{2}=\frac{\left(  \vec{k}%
+\vec{q}\right)  ^{2}}{2\beta \left(  m+M\right)  }$):%
\begin{align}
&  \int_{0}^{\infty}dt\sin \left(  \omega t\right)  \left \{  \left[  1+n\left(
\omega_{\vec{q}}\right)  \right]  e^{-C^{2}t^{2}+iA^{+}t}+n\left(
\omega_{\vec{q}}\right)  e^{-C^{2}t^{2}+iA^{-}t}\right \} \nonumber \\
&  =\frac{1}{2iC}\left \{  \left[  1+n\left(  \omega_{\vec{q}}\right)  \right]
\left[  \frac{\sqrt{\pi}}{2}e^{-\frac{\left(  A^{+}+\omega \right)  ^{2}%
}{4C^{2}}}+iF\left(  \frac{A^{+}+\omega}{2C}\right)  -\frac{\sqrt{\pi}}%
{2}e^{-\frac{\left(  A^{+}-\omega \right)  ^{2}}{4C^{2}}}-iF\left(  \frac
{A^{+}-\omega}{2C}\right)  \right]  \right. \nonumber \\
&  \left.  +n\left(  \omega_{\vec{q}}\right)  \left[  \frac{\sqrt{\pi}}%
{2}e^{-\frac{\left(  A^{-}+\omega \right)  ^{2}}{4C^{2}}}+iF\left(  \frac
{A^{-}+\omega}{2C}\right)  -\frac{\sqrt{\pi}}{2}e^{-\frac{\left(  A^{-}%
-\omega \right)  ^{2}}{4C^{2}}}-iF\left(  \frac{A^{-}-\omega}{2C}\right)
\right]  \right \}  .
\end{align}
Where we introduced the Dawson integral $F\left(  x\right)  $:
\begin{align}
F\left(  x\right)   &  =e^{-x^{2}}\int_{0}^{x}e^{y^{2}}dy\nonumber \\
&  =\frac{1}{2}\sqrt{\pi}e^{-x^{2}}\text{erfi}\left(  x\right)  ,
\end{align}
where erfi$\left(  x\right)  $ is the imaginary error function: erfi$\left(
x\right)  =-i\operatorname{erf}\left(  ix\right)  $, with $\operatorname{erf}%
\left(  x\right)  $ the error function. This finally results in the following
expression for the imaginary part of the self energy:
\begin{align}
\operatorname{Im}\left[  \Sigma \left(  \omega,\vec{k}\right)  \right]   &
=\frac{\sqrt{2\pi \beta \left(  m+M\right)  }}{2m_{I}N\hbar}\sum_{n,n^{\prime
}=0}^{\infty}\sum_{\vec{q}\neq0}\left \vert V_{\vec{q}}\right \vert ^{2}%
\frac{\left(  \vec{k}.\vec{q}\right)  ^{2}}{k^{2}}\left(  \vec{k}+\vec
{q}\right)  ^{2\left(  n+n^{\prime}\right)  -1}B\left(  \beta,n,n^{\prime
}\right) \nonumber \\
&  \times e^{-a^{2}\left(  \beta \right)  \left(  \vec{k}+\vec{q}\right)  ^{2}%
}\left \{  \left[  1+n\left(  \omega_{\vec{q}}\right)  \right]  \left[
e^{-\frac{\beta \left(  m+M\right)  \left(  A^{+}+\omega \right)  ^{2}}{2\left(
\vec{k}+\vec{q}\right)  ^{2}}}-e^{-\frac{\beta \left(  m+M\right)  \left(
A^{+}-\omega \right)  ^{2}}{2\left(  \vec{k}+\vec{q}\right)  ^{2}}}\right]
\right. \nonumber \\
&  \left.  +n\left(  \omega_{\vec{q}}\right)  \left[  e^{-\frac{\beta \left(
m+M\right)  \left(  A^{-}+\omega \right)  ^{2}}{2\left(  \vec{k}+\vec
{q}\right)  ^{2}}}-e^{-\frac{\beta \left(  m+M\right)  \left(  A^{-}%
-\omega \right)  ^{2}}{2\left(  \vec{k}+\vec{q}\right)  ^{2}}}\right]
\right \}  . \label{ImagZelf}%
\end{align}
For the real part of (\ref{Zelf2}) we have:%
\begin{align}
\operatorname{Re}\left[  \Sigma \left(  \omega,\vec{k}\right)  \right]   &
=\frac{2}{m_{I}N\hbar}\sum_{n,n^{\prime}=0}^{\infty}\sum_{\vec{q}\neq
0}\left \vert V_{\vec{q}}\right \vert ^{2}\frac{\left(  \vec{k}.\vec{q}\right)
^{2}}{k^{2}}\left(  \vec{k}+\vec{q}\right)  ^{2\left(  n+n^{\prime}\right)
}B\left(  \beta,n,n^{\prime}\right)  e^{-a^{2}\left(  \beta \right)  \left(
\vec{k}+\vec{q}\right)  ^{2}}\nonumber \\
&  \times \operatorname{Im}\left(  \int_{0}^{\infty}dt\left[  1-\cos \left(
\omega t\right)  \right]  \left \{  \left[  1+n\left(  \omega_{\vec{q}}\right)
\right]  e^{-\frac{\left(  \vec{k}+\vec{q}\right)  ^{2}}{2\beta \left(
m+M\right)  }t^{2}+iA^{+}t}\right.  \right. \nonumber \\
&  \left.  \left.  +n\left(  \omega_{\vec{q}}\right)  e^{-\frac{\left(
\vec{k}+\vec{q}\right)  ^{2}}{2\beta \left(  m+M\right)  }t^{2}+iA^{-}%
t}\right \}  \right)  .
\end{align}
The time-integration is in this case:%
\begin{align}
&  \int_{0}^{\infty}dt\left[  1-\cos \left(  \omega t\right)  \right]  \left \{
\left[  1+n\left(  \omega_{\vec{q}}\right)  \right]  e^{-C^{2}t^{2}+iA^{+}%
t}+n\left(  \omega_{\vec{q}}\right)  e^{-C^{2}t^{2}+iA^{-}t}\right \}
\nonumber \\
&  =\frac{1}{2C}\left \{  \left[  1+n\left(  \omega_{\vec{q}}\right)  \right]
\left[  \sqrt{\pi}e^{-\frac{\left(  A^{+}\right)  ^{2}}{4C^{2}}}+2iF\left(
\frac{A^{+}+\omega}{2C}\right)  -\frac{\sqrt{\pi}}{2}e^{-\frac{\left(
A^{+}+\omega \right)  ^{2}}{4C^{2}}}-iF\left(  \frac{A^{+}+\omega}{2C}\right)
\right.  \right. \nonumber \\
&  \left.  -\frac{\sqrt{\pi}}{2}e^{-\frac{\left(  A^{+}+\omega \right)  ^{2}%
}{4C^{2}}}-iF\left(  \frac{A^{+}-\omega}{2C}\right)  \right]  +n\left(
\omega_{\vec{q}}\right)  \left[  \sqrt{\pi}e^{-\frac{\left(  A^{-}\right)
^{2}}{4C^{2}}}+2iF\left(  \frac{A^{-}+\omega}{2C}\right)  \right. \nonumber \\
&  \left.  \left.  -\frac{\sqrt{\pi}}{2}e^{-\frac{\left(  A^{-}+\omega \right)
^{2}}{4C^{2}}}-iF\left(  \frac{A^{-}+\omega}{2C}\right)  -\frac{\sqrt{\pi}}%
{2}e^{-\frac{\left(  A^{-}+\omega \right)  ^{2}}{4C^{2}}}-iF\left(  \frac
{A^{-}-\omega}{2C}\right)  \right]  \right \}  .
\end{align}
This results in the following expression for the real part of the self
energy:
\begin{align}
\operatorname{Re}\left[  \Sigma \left(  \omega,\vec{k}\right)  \right]   &
=\frac{\sqrt{2\beta \left(  m+M\right)  }}{m_{I}N\hbar}\sum_{n,n^{\prime}%
=0}^{\infty}\sum_{\vec{q}\neq0}\left \vert V_{\vec{q}}\right \vert ^{2}%
\frac{\left(  \vec{k}.\vec{q}\right)  ^{2}}{k^{2}}\left(  \vec{k}+\vec
{q}\right)  ^{2\left(  n+n^{\prime}\right)  -1}B\left(  \beta,n,n^{\prime
}\right) \nonumber \\
&  e^{-a^{2}\left(  \beta \right)  \left(  \vec{k}+\vec{q}\right)  ^{2}}\left(
\left(  1+n\left(  \omega_{\vec{q}}\right)  \right)  \left \{  2F\left[
\frac{\sqrt{2\beta \left(  m+M\right)  }A^{+}}{2\left \vert \vec{k}+\vec
{q}\right \vert }\right]  \right.  \right. \nonumber \\
&  \left.  -F\left[  \frac{\sqrt{2\beta \left(  m+M\right)  }\left(
A^{+}+\omega \right)  }{2\left \vert \vec{k}+\vec{q}\right \vert }\right]
-F\left[  \frac{\sqrt{2\beta \left(  m+M\right)  }\left(  A^{+}-\omega \right)
}{2\left \vert \vec{k}+\vec{q}\right \vert }\right]  \right \} \nonumber \\
&  +n\left(  \omega_{\vec{q}}\right)  \left \{  2F\left[  \frac{\sqrt
{2\beta \left(  m+M\right)  }A^{-}}{2\left \vert \vec{k}+\vec{q}\right \vert
}\right]  -F\left[  \frac{\sqrt{2\beta \left(  m+M\right)  }\left(
A^{-}+\omega \right)  }{2\left \vert \vec{k}+\vec{q}\right \vert }\right]
\right. \nonumber \\
&  \left.  \left.  -F\left[  \frac{\sqrt{2\beta \left(  m+M\right)  }\left(
A^{-}-\omega \right)  }{2\left \vert \vec{k}+\vec{q}\right \vert }\right]
\right \}  \right)  . \label{ReZelf}%
\end{align}

\bibliographystyle{phpf}
\bibliography{gereduceerdeDimensies}

\end{document}